\documentclass[epj,nopacs,superscriptaddress,longbibliography]{svjour}
\usepackage{graphics}
\usepackage{amsmath,amssymb}
\usepackage{cite}

\usepackage[breaklinks,colorlinks,citecolor=blue,linkcolor=blue,urlcolor=blue]{hyperref}
\begin{document}


%
\title{\small \hspace{12cm}\href{https://doi.org/10.1051/epjap/2023230060}{Eur. Phys. J. Appl. Phys. {\bf 98}, 39 (2023)}\\

\Large
Atomic relaxation and electronic structure in twisted bilayer MoS$_2$ with rotation angle of 5.09 degrees}
\author{Somepalli Venkateswarlu,\inst{1}
Ahmed Misssaoui,\inst{1} 
Andreas Honecker,\inst{1} \and 
Guy Trambly de Laissardi\`ere\inst{1}
}                     
%
%
\institute{\inst{1}Laboratoire de Physique Th\'eorique et Mod\'elisation, CY Cergy Paris Universit\'e, CNRS, 95302 Cergy-Pontoise, France }
%
\date{June 2023}

%
\abstract{
It is now well established theoretically and experimentally that a
moir\'e pattern, due to a rotation of two atomic layers with respect to each other, creates low-energy flat bands.
First discovered in twisted bilayer graphene, these new electronic states are at the origin of strong electronic correlations and even of unconventional superconductivity. 
Twisted bilayers (tb) of transition metal dichalcogenides 
(TMDs) also exhibit flat bands around their semiconductor gap at small rotation angles. 
In this paper, we present a DFT study to analyze the effect of the atomic relaxation on the low-energy bands of tb-MoS$_2$ with a rotation angle of 5.09$^\circ$. 
We show that in-plane atomic relaxation is not essential here, while out-of-plane relaxation dominates the electronic structure.
We propose a simple and efficient atomic model to predict this relaxation.
\PACS{
      {PACS-key}{discribing text of that key}   \and
      {PACS-key}{discribing text of that key}
     } 
} 
\authorrunning{S. Venkateswarlu {\it et al.}}
\titlerunning{DFT study of atomic relaxation in $5.09^\circ$ twisted bilayer MoS$_2$}

\maketitle

\section{Introduction}
\label{intro}

The broad family of transition metal dichalcogenides
(TMDs) 
\cite{Wang15_reviewTMD,Liu15_reviewTMD,Duong17_review} 
offers the possibility to stack two layers with a small angle of rotation $\theta$ to each other, thus forming a moir\'e pattern superstructure. 
These twisted bilayers have given rise to numerous experimental 
\cite{vanderZande14,Liu14,Huang14,Huang16,Zhang17,Trainer17,Yeh16,Lin18,Pan18,Zhang20_tTMDC} 
and theoretical 
\cite{Roldan14b,Fang15,Cao15,Wang15,Constantinescu15,Tan16,Lu17,Naik18,
Conte19,Maity19,Tang20,Wu20,Lu20,Pan20,Venky20,
Zhan20,Zhang20,Naik20,Xian21,Vitale21,Angeli21,He21} studies to understand electronic states that are confined  by a moir\'e pattern  in semiconductor materials. 
Many  of these studies analyze the interlayer distances, 
the possible atomic  relaxation, 
the transition from a direct band gap in the monolayer system 
to an indirect band gap in bilayer systems, 
and more generally the effect of interlayer coupling 
in these twisted bilayer 2D systems at various rotation 
angles $\theta$. 
For small values of $\theta$, the emergence of flat bands 
has been established  from first-principles density 
functional theory (DFT) calculations \cite{Naik18,Naik20} and tight-binding (TB) calculations \cite{Venky20,Zhan20,Zhang20} in twisted bilayer MoS$_2$ (tb-MoS$_2$), 
and observed in a 3$^\circ$ twisted bilayer WSe$_2$ sample by using 
scanning tunneling spectroscopy \cite{Zhang20_tTMDC}. 
It has been also
shown numerically \cite{Lu20} that Lithium intercalation in tb-MoS$_2$ 
increases interlayer coupling and thus promotes flat bands around the gap. 
There is also experimental evidence that moir\'e patterns may give rise to 
confined states due to the mismatch of the lattice parameters in 
MoS$_2$--WSe$_2$ heterobilayers \cite{Pan18}.

Most theoretical investigations of the electronic structure of bilayer 
MoS$_2$ are density-functional theory (DFT) studies 
\cite{vanderZande14,Liu14,Huang14,Zhang17,Roldan14b,Fang15,Cao15,Wang15,Constantinescu15,Huang16,Tan16,Lu17,Trainer17,Naik18,Naik20,Debbichi14,Tao14,He14,sun2020effects} 
with eventually a Wannier wave function analysis \cite{Fang15}. 
To provide systematic 
analysis as a function of the rotation angle 
$\theta$, in particular for small angles, {\it i.e.}, very large moir\'e pattern
cells for which DFT calculations are not feasible, several TB models, based on Slater-Koster (SK) parameters \cite{Slater54}, have 
been proposed for monolayer MoS$_2$ 
\cite{Cappelluti13,Rostami13,Zahid13,Ridolfi15,SilvaGuillen16} and 
multi-layer MoS$_2$ \cite{Cappelluti13,Roldan14b,Fang15,Zahid13,
Zhan20,Zhang20}.
Following these efforts, 
we have proposed \cite{Venky20} a SK-TB set of parameters for non-relaxed structures,
{\it i.e.}, rigidly twisted bilayers,
that match correctly the DFT bands around the gap of tb-MoS$_2$ with 
rotation angles $\theta > 7^\circ$. This SK-TB model, with the same 
parameters, is then used for smaller angles in order to describe the 
states confined by the moir\'e pattern. 
For 
$\theta < \theta_C \simeq 5^\circ$,
the valence band with the highest energy is separated from the other 
valence states by a minigap of a few meV. 
In addition, the width of this 
band decreases as $\theta$ decreases so that the average velocity of these 
electronic states reaches 0 for $\theta \lesssim 2^\circ$ such that almost flat bands emerge at these angles. 
This is reminiscent of the vanishing of the 
velocity at certain ``magic'' rotation angles in twisted bilayer graphene \cite{Trambly10,SuarezMorell10,Bistritzer11}. 
However, in  bilayer MoS$_2$ it arises for an interval of angles and not a set of specific values. 
Other minigaps and flat bands are also found in the conduction 
band. The confined states that are closest to the gap are localized in the 
AA stacking regions of the moir\'e pattern, as in twisted bilayer  graphene.
However, for small angles, it has been shown \cite{Naik18,Naik20,Vitale21} that atomic relaxation strongly modifies these low-energy bands, in particular their approximate degeneracy.
A better understanding of atomic relaxation and its effects on the electronic structure is therefore essential not only for very small angles but also for larger ones ($\theta \simeq 5^\circ$).

\begin{figure}[t!]
\begin{center}
\resizebox{0.5\textwidth}{!}{%
  ~~~~~\includegraphics{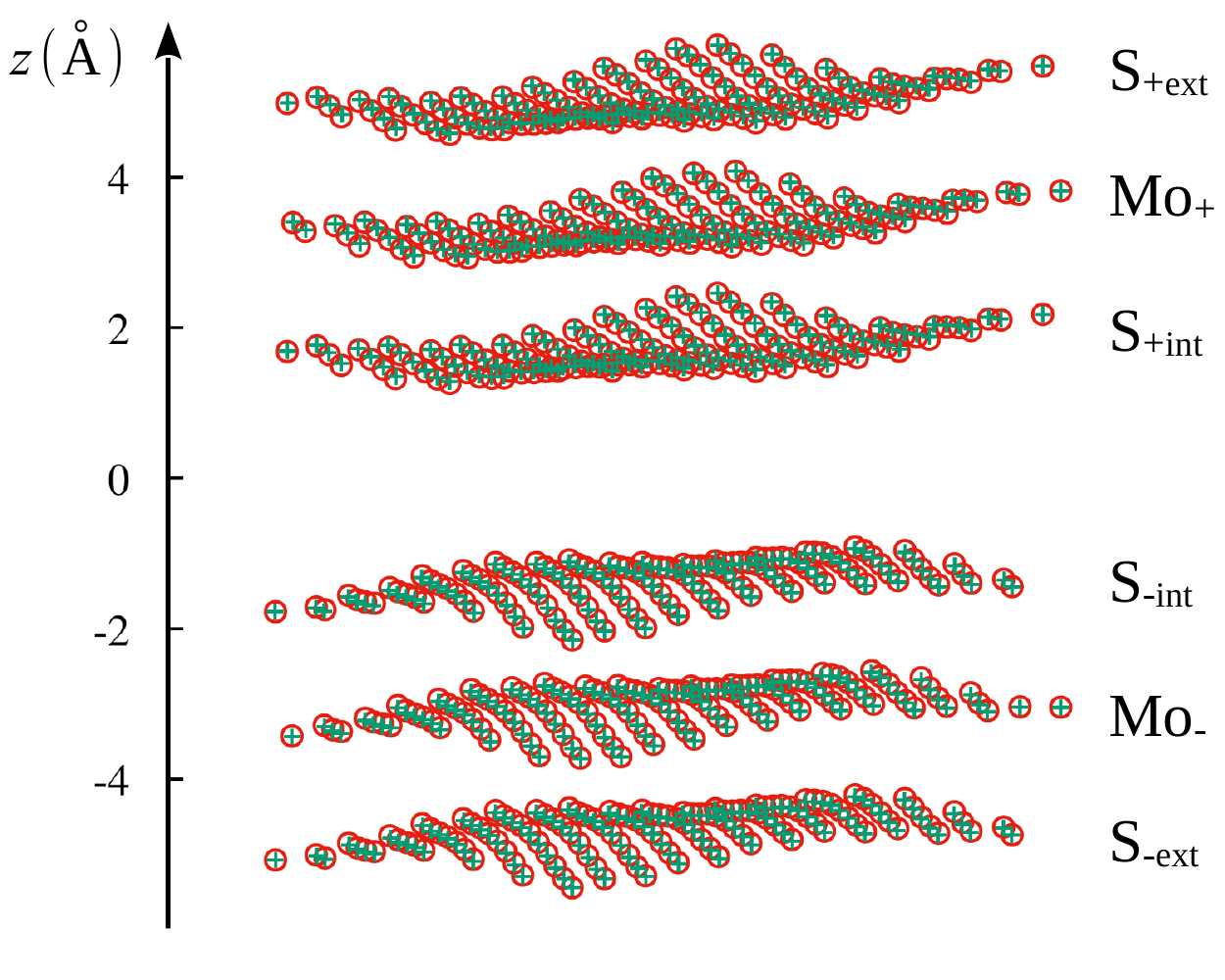}
}

\vspace{0.1cm}

\resizebox{0.45\textwidth}{!}{%
  \includegraphics{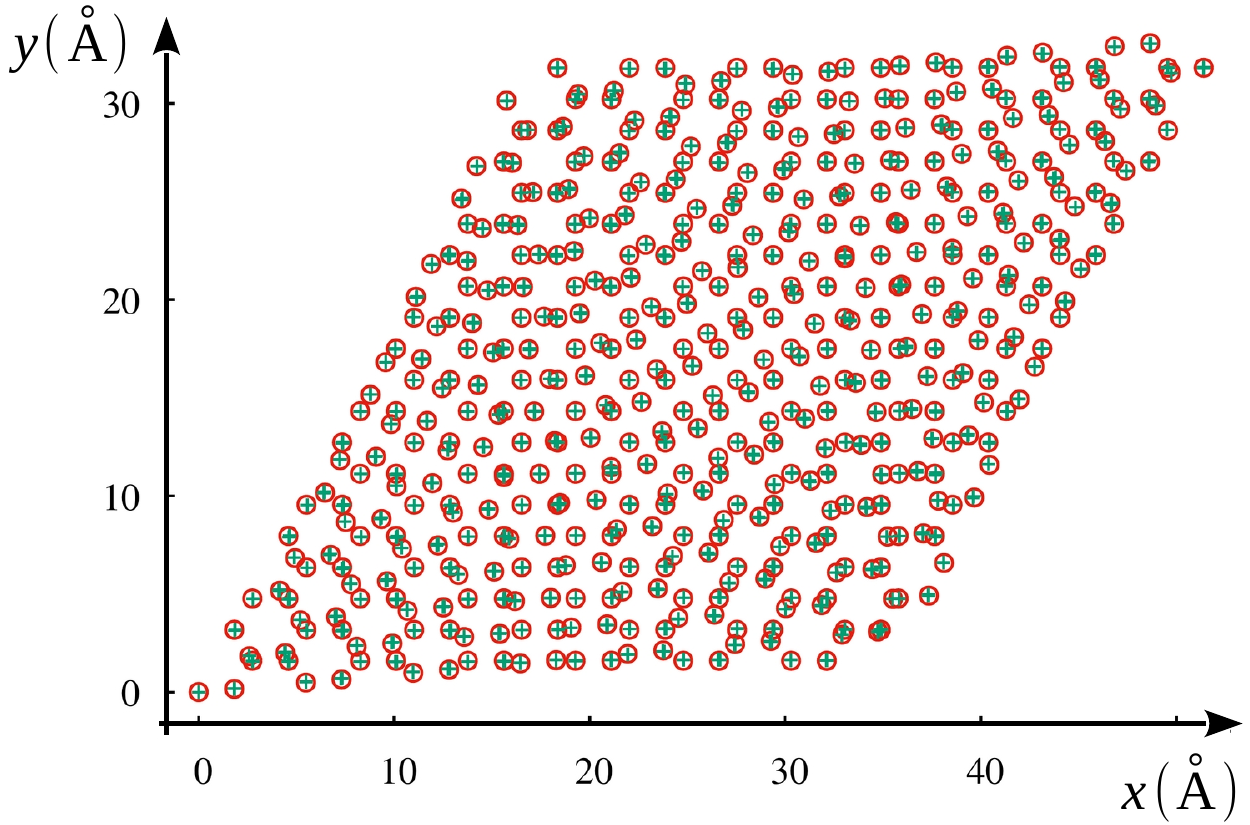}
}
\end{center}
\caption{Sketch of DFT-relaxed tb-MoS$_2$ with rotation angle $\theta = 5.09^\circ$. 
All atoms in a moir\'e cell are represented: 
Red circles are DFT calculation and green crosses are the positions calculated with $z$-modulation only (Eq.~(\ref{Eq_zMod})).  
(top) Perspective side view of a moir\'e cell (layer ``$+$’’ (``$-$'') is located in the $z>0$ ($z<0$) region), and  (bottom) the corresponding top view.
}
\label{fig_struct}      
\end{figure}

In this paper, we present a DFT study of tb-MoS$_2$ with a rotation angle of $\theta = 5.09^\circ$.
This angle is approximately the angle below which the highest energy valence band (just below the gap) is isolated from the rest of the valence states by a minigap for non-relaxed structures.  
We show that, unlike the case of very small angles \cite{Naik18,Naik20,Vitale21}, atomic relaxation amounts to essentially out-of-plane atomic displacements along the $z$-direction that can be simulated simply as a function of the atomic positions in the moir\'e cell. 
This simple atomic model allows to understand the origin of the modifications of the electronic structure induced by the relaxation.

\section{Structure and DFT calculations}

The construction of the commensurable twisted bilayer we study  is explained in detail in Refs.~\cite{Venky20,Venky_these}. It corresponds to $(n,m)=(6,7)$ with 762 atoms per unit cell. 
Starting from an AA stacked bilayer (where Mo atoms of a layer lie above a Mo atom of the other layer, and S atoms of a layer lie above an S atom of the other layer), one layer (layer ``$+$'') is rotated with respect to the other layer (layer ``$-$'') by the angle $\theta = 5.09^\circ$  around an axis containing two Mo atoms. 
Thus, AA stacking regions are located at the corner of the moir\'e cell (Fig.~\ref{fig_struct}(bottom)).
BA’ stacking regions (where S$-$ atoms lie above a Mo$+$, and Mo$-$ (S$+$) do not lie above any atom of layer $+$ (layer $-$)), and 
AB’ stacking regions (regions where Mo$-$ lie above an S$+$, and S$-$ (Mo$+$) do not lie above any atom
of layer $+$ (layer $-$))  are located at
$1/3$ and $2/3$ of the long diagonal of the moir\'e cell, respectively.

The DFT 
calculations were carried out with the ABINIT software 
\cite{Gonze02,Gonze09,Gonze16}. We have checked previously 
\cite{Venky20,Venky_these} that LDA \cite{Jones89} and GGA \cite{Perdew96} 
+ Van der Waals exchange-cor\-re\-la\-tion functionals yield very similar 
results, so all the results presented here are based on LDA calculations, 
which require less computation time for large systems. While the results 
presented here may not be quantitatively very accurate, they are 
sufficient to qualitatively show the importance of out-of-plane atomic 
displacements on the electronic structure of the low-energy flat bands. 
Likewise, the exact value of the band gap may not be very precise in LDA, 
but the evolution of this gap and the existence of minigaps are 
significant. The Brillouin zone was sampled by a k-point mesh in 
reciprocal space within the Monkhorst-Pack scheme \cite{Monkhorst76}. One 
k-point is used for atomic relaxation and a 2$\times$2 k-grid for the 
self-consistency procedure of the electronic structure calculation. The 
kinetic energy cutoff was chosen to be 408\,eV. We checked that a 3$\times$3 
k-grid for the self-consistency procedure and an energy cutoff of 544.22\,eV 
yield very similar bands for the relaxed structure. The structural 
optimization of atomic positions is done by using the 
Broyden-Fletcher-Goldfarb-Shanno minimization. A vacuum region of 
10\,\AA{} was inserted between the MoS$_2$ bilayers to avoid spurious 
interactions between periodic images. In our calculations, the spin-orbit 
coupling (SOC) is not taken into account in order to reduce the 
calculation time. SOC is important for TMDs, as it introduces some band 
splittings close to the gap 
\cite{Ridolfi15,SilvaGuillen16,Naik18,Zhan20,He21}, however it will not 
change the existence or not of low-energy flat bands in tb-MoS$_2$, so it 
is not essential for the present study.

\section{Atomic relaxation}

\begin{figure}[t!]
\begin{center}
\resizebox{0.499\textwidth}{!}{%
  ~~~~~\includegraphics{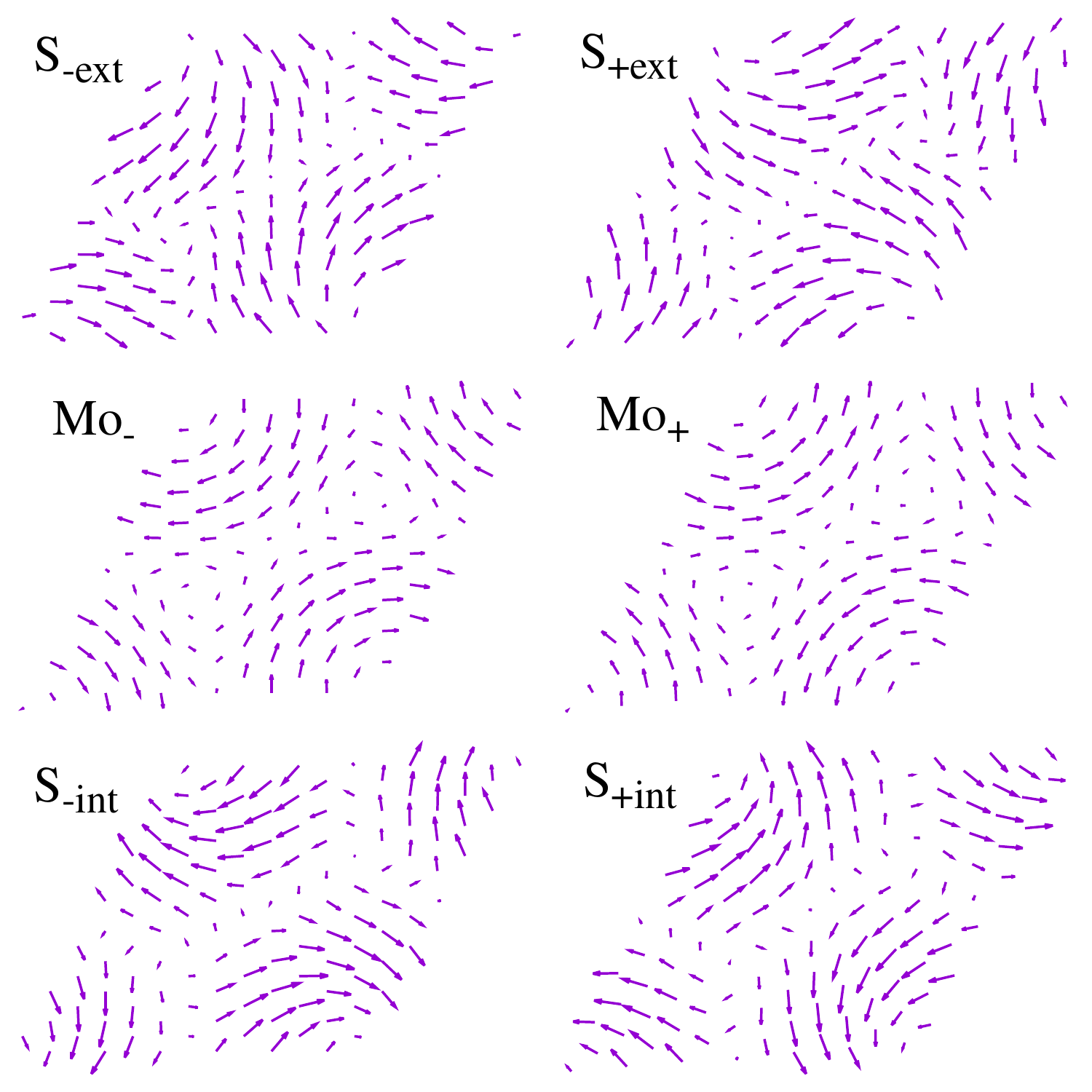}
}
\end{center}
\caption{In-plane displacement vectors of the atoms in the moir\'e cell of the DFT-relaxed structure with respect to the rigidly twisted bilayer (Non-relaxed structure). 
To be visible, 
the displacement vectors $\vec{\tau}_i$ of each atom $i$ have been multiplied by $40$.
}
\label{fig_struct_inPlane}      
\end{figure}

The DFT-relaxed atomic structure of tb-MoS$_2$ with rotation angle $\theta = 5.09^\circ$ is shown in Fig.~\ref{fig_struct}.
Remarkably, the in-plane displacement ${\vec \tau}_i$ of each atom $i$ with respect to the rigidly twisted structure is rather small.  
Indeed, the average $\| {\vec \tau}_i \|$ is 
$0.04 \pm 0.02$\,\AA{}, 
$0.03 \pm 0.02$\,\AA{}, and
$0.05 \pm 0.02$\,\AA{}, 
for atoms S$_{\pm{\rm ext}}$,
Mo$_{\pm}$, and
S$_{\pm{\rm int}}$, respectively. 
Such displacements are almost not visible in Fig.~\ref{fig_struct}(bottom), and they have little effect on the electronic structure (see next section).
This result shows that the strong in-plane displacements  obtained for the smallest angles \cite{Naik18,Naik20,Vitale21} are not too important for $\theta \simeq 5^\circ$.
However, it is interesting to note that these small displacements are precursors to the larger displacements and shear solitons obtained for the smallest angles \cite{Naik18}, as shown in Fig.~\ref{fig_struct_inPlane}.
These in-plane displacements  tend to reduce the AA stacking regions with respect to the AB stacking regions to minimize the energy \cite{Naik18,Naik20,Vitale21}.

\begin{table}[tb!]
\caption{Atomic positions: Average $\bar{z}$ value per layer and the 
corresponding standard deviation $\sigma_z$, for layer $+$ ($z>0$) and 
layer $-$ ($z<0$)  in non-relaxed, relaxed, and $z$-mod tb-MoS$_2$ 
structures. The atoms of the $z$-mod structure have the same in-plane $xy$ 
coordinates as the non-relaxed structure and their $z$ coordinates are 
calculated from Eq.\,(\ref{Eq_zMod}) with $z_{\rm AA}$ and $\rm z_{AB}$ 
({\it i.e.}, in AB' and BA' stacking regions) of the DFT-relaxed 
structure. For each layer $\delta z = |z_{\rm AA} - z_{\rm AB}|$.  The four 
last lines correspond to simple AA and AB' (BA') bilayer stackings for 
which $\bar{z} = z_{\rm AA}$ and $z_{\rm AB}$, respectively. All 
calculations were performed using the LDA pseudopotential, except $z_{\rm 
AA}^{\rm \footnotesize{GGA}}$ and $z_{\rm AB}^{\rm \scriptsize{GGA}}$ that 
are calculated with the GGA PBE pseudopotential \cite{Perdew96} including 
Van der Waals corrections (D2) \cite{Grimme06,Gonze16}. The latter calculations 
show that the displacements obtained by the LDA are qualitatively correct. 
The $\bar{z}$ values for the two layers are symmetric with respect to 
$z=0$. Distances are in \AA.
\label{Tableau}}
\begin{tabular}{lllll}
\hline\noalign{\smallskip}
Structure    &  & S$_{\rm \pm ext}$ & Mo$_{\rm \pm}$ & S$_{\rm \pm int}$ \\
\noalign{\smallskip}\hline\noalign{\smallskip}
non-relaxed & $\bar{z}$      & $\pm 4.96$ & $\pm 3.40$ & $\pm 1.84$ \\
relaxed     & $\bar{z}$      & $\pm 4.86$ & $\pm 3.21$ & $\pm 1.56$ \\
            & $\sigma_z$     & $0.10$     & $0.10$     & $0.10$ \\
            & $z_{\rm AA}$   & $\pm 5.08$ & $\pm 3.43$ & $\pm 1.78$ \\
            & $z_{\rm AB}$   & $\pm 4.72$ & $\pm 3.07$ & $\pm 1.42$ \\
            & $\delta z$     & $0.36$     & $ 0.36$    & $ 0.36$ \\
$z$-mod     & $\bar{z}$      & $\pm 4.84$ & $\pm 3.19$ & $\pm 1.54$ \\
            & $\sigma_z$     & $0.10$     & $0.10$     & $0.10$ \\
bilayer AA  & $z_{\rm AA}$  & $\pm 5.11$    & $\pm 3.50$ & $\pm 1.88$ \\
  & $z_{\rm AA}^{\rm \footnotesize{GGA}}$  & $\pm 4.99$    & $\pm 3.42$ & $\pm 1.86$ \\
bilayer AB' (BA') & $z_{\rm AB}$  & $\pm 4.62$    & $\pm 3.01$ & $\pm 1.39$ \\
  & $z_{\rm AB}^{\rm \scriptsize{GGA}}$  & $\pm 4.67 $    & $\pm 3.11 $ & $\pm 1.55 $ \\
\noalign{\smallskip}\hline
\end{tabular}
\end{table}

On the other hand, the displacements along the $z$-direction are important. 
For each atomic layer, the mean values of the $z$-coordinate and the corresponding standard deviation are given in Table\,\ref{Tableau}.
As expected given the interplane distances in the simple stacking cases  (see Ref.\,\cite{He14} and Table\,\ref{Tableau}), the distance between layers is greater in the AA stacking regions than in AB' (BA') stacking regions. 
Inspired by the work of Koshino {\it et al.}\ \cite{Koshino18} for twisted bilayer graphene, we propose the following atomic model (``$z$-mod'' model) where $xy$ in-plane atomic coordinates are those of the rigidly twisted bilayer and the modulation of the atomic $z$-coordinates are calculated by
\begin{equation}
z(\vec r) = \frac{(z_{\rm AA} +2 z_{\rm AB})}{3}  
+  \frac{(z_{\rm AA} - z_{\rm AB})}{9} 
\sum_{i=1}^6 {\rm cos} \left( {\vec G}_i \cdot \vec r \right),
\label{Eq_zMod}
\end{equation}
where $\vec r$ are the non-relaxed positions of the atoms in the  rigidly twisted moir\'e cell, and ${\vec G}_i$ are the six vectors of the reciprocal lattice that define the first Brillouin zone of the moir\'e pattern. 
For each atomic layer, {\it i.e.}, for the six Mo and S atomic layers of tb-MoS$_2$, $z_{\rm AA}$ and $z_{\rm AB}$ are the $z$ values of the DFT-relaxed structure at $\vec r = \vec 0$ (AA stacking region)  and   $\vec r$ at AB' or BA' stacking regions 
(Table \ref{Tableau}).  
Figure~\ref{fig_struct} shows that DFT-relaxed positions and $z$-mod positions fit well together.

\section{Electronic band dispersion}

\begin{figure}
\begin{center}
\resizebox{0.45\textwidth}{!}{%
  \includegraphics{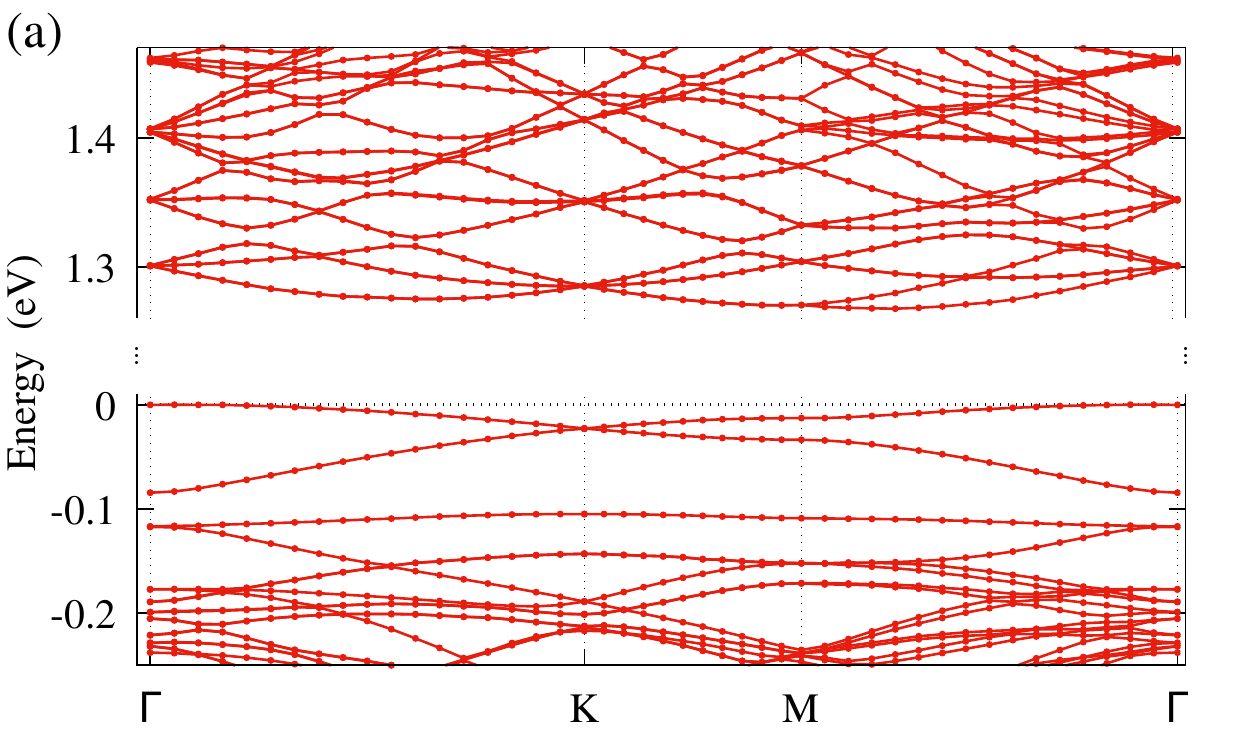}
}

\resizebox{0.45\textwidth}{!}{%
  \includegraphics{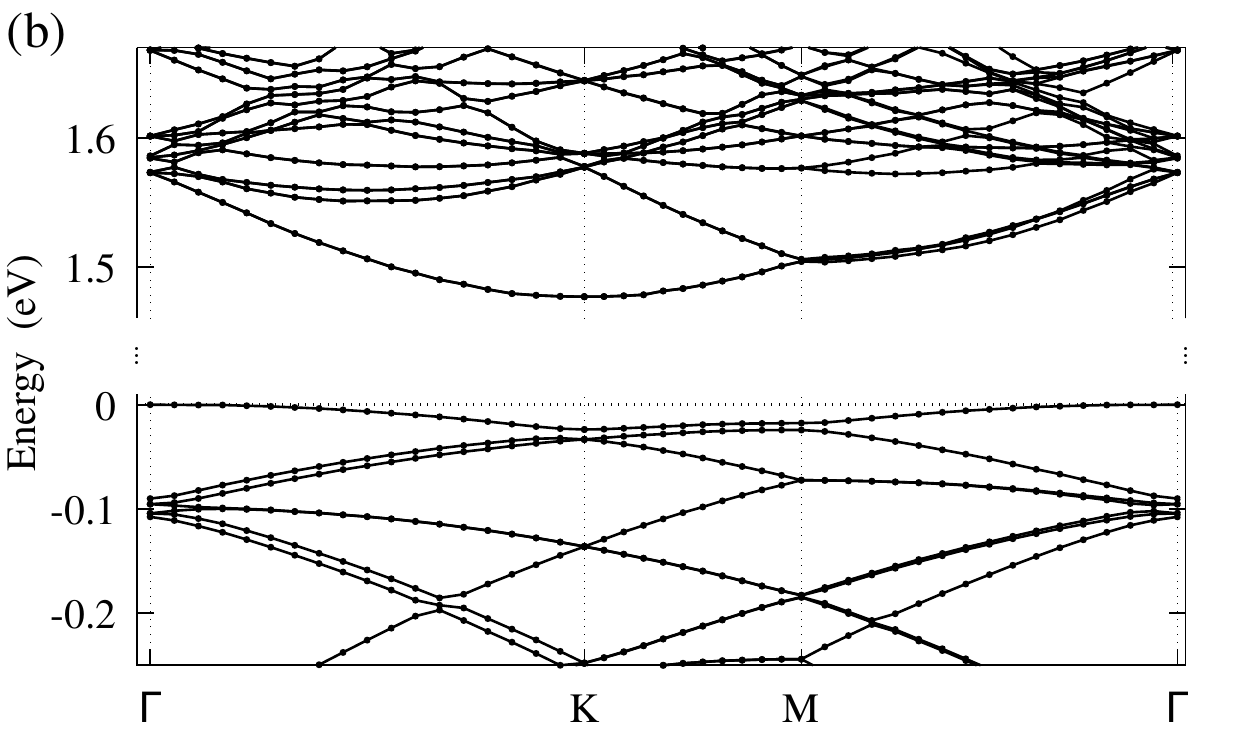}
}

\resizebox{0.45\textwidth}{!}{%
  \includegraphics{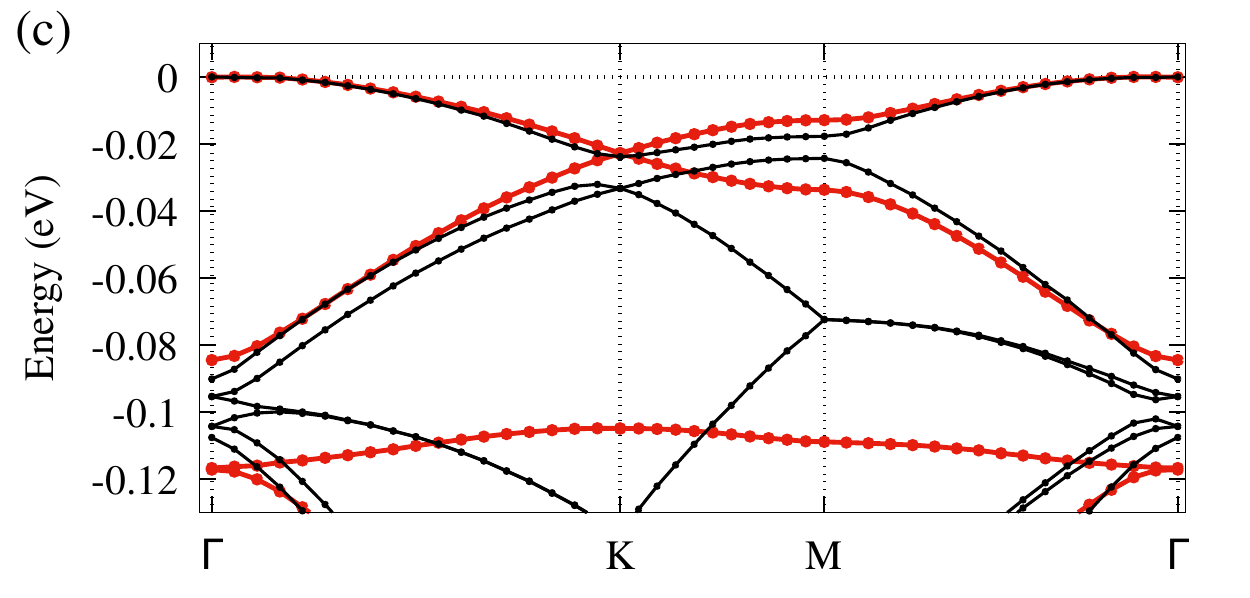}
}

\end{center}
\caption{DFT bands around the gap for (a) relaxed and (b) non-relaxed tb-MoS$_2$ with $\theta = 5.09^\circ$. 
(c) Zoom in on the bands just below the gap for (red) relaxed and (black) non-relaxed structures.
The energy $0$ is fixed at the maximum energy of the valence band. Lines are guides for the eye. }
\label{fig_bands_R_NR}      
\end{figure}

The DFT band dispersions are shown for the DFT-relaxed bilayer and the rigidly twisted bilayer (non-relaxed bilayer) in Fig.~\ref{fig_bands_R_NR}(a) and \ref{fig_bands_R_NR}(b),
respectively. 
For the non-relaxed bilayer the interlayer distance is that obtained for simple AA stacking, like in our previous calculations \cite{Venky20}.
In the non-relaxed tb-MoS$_2$, the minimum of the conduction band is at K like for monolayer MoS$_2$, which is no longer the case after atomic relaxation. 
Fig.\,\ref{fig_bands_R_NR}(c) shows a zoom of the highest-energy valence bands. 
The relaxation does not change the valence band maximum energy at K. 
However it leads to significant modifications of the bands close to the gap. 
For the non-relaxed structure \cite{Venky20}, when $\theta < \theta_c$, the highest-energy band is non-degenerate and isolated from the other valence bands by a minigap. 
For $\theta = 5.09^\circ$, this minigap is equal to $\sim 0.5$\,meV (Fig.~\ref{fig_bands_R_NR}(c)).  
For the relaxed structure there is no isolated single valance band, but 2 bands which cross in K while remaining linear in $k$ (around K). 
This result is similar to that of Naik {\it et al.}\ \cite{Naik18,Naik20} and Vitale {\it et al.}\ \cite{Vitale21}, 
obtained by using a  multiscale approach, a pair potential for relaxation, and DFT or TB calculations for the band dispersion.
There are nevertheless small differences.
For instance,
in our calculation, the bandwidth for these first two valence bands is 84\,meV with respect to $\sim 120$\,meV in Ref.~\cite{Naik20}. 
Unlike that previous calculation, these two bands are isolated from the rest of the valence bands by a minigap of $\sim 20$\,meV.

\begin{figure}
\begin{center}
\resizebox{0.45\textwidth}{!}{%
  \includegraphics{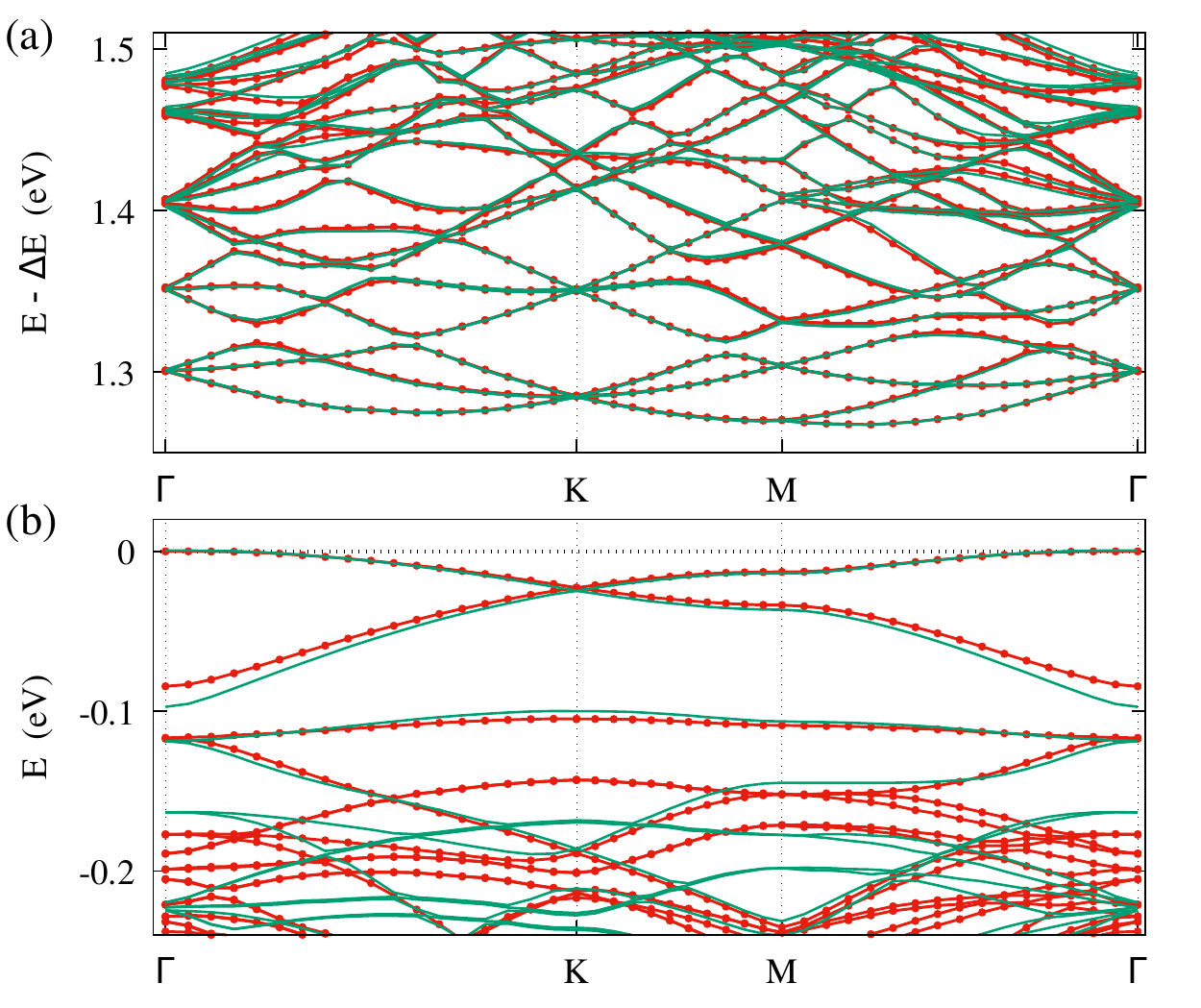}
}
\end{center}
\caption{DFT bands around the gap for (red) DFT-relaxed tb-MoS$_2$ with $\theta = 5.09^\circ$, and (green) the structure with $z$-modulation calculated by the equation (\ref{Eq_zMod}):
(a) conduction bands,
(b) valence bands. 
The energy $0$ is fixed at the maximum energy of the valence bands. 
As these two structures do not have the same gap, an energy shift $\Delta E$ is introduced to compare conduction bands easily: 
$\Delta E ({\rm DFT~relaxed}) = 0$, and 
$\Delta E (z$-mod$) = -36$\,meV.
Lines are guides for the eye.}
\label{fig_bands_R_z-mod}      
\end{figure}

To test the validity of the structure with the $z$-mod\-u\-la\-tion only 
(Eq.~(\ref{Eq_zMod})), 
bands around the gap of DFT-relaxed and $z$-mod structures are compared in
Fig.~\ref{fig_bands_R_z-mod}. 
The gaps of the two structures are slightly different (difference of 36\,meV), which may be due to the difference between the average $z$ values per atomic layer (Table \ref{Tableau}).
The conduction bands are nevertheless almost the same and valence bands are very similar. 
In particular, the two valence bands closest to the gap that are characteristic of the atomic relaxation effect are well reproduced. 
Therefore, the simplified relaxation model given by formula (\ref{Eq_zMod}) is sufficient to account for the low-energy flat bands due to a moir\'e pattern at not too small rotation angles. 

\section{Conclusion}

We have performed a DFT study of the atomic relaxation and the electronic band dispersion of twisted bilayer MoS$_2$ with a rotation angle equal to $5.09^\circ$.
Contrary to what has been observed for very small angles (typically less than $\sim 2^\circ$) 
\cite{Naik18,Naik20,Vitale21}, 
the in-plane atomic displacements with respect to a rigidly twisted bilayer are very small, and they have almost no effect on band dispersion.  
However, the out-of-plane displacements are large and significantly alter the low-energy bands around the gap. 
These atomic displacements can be modeled by a simple formula that depends only on the interlayer distances in the AA and AB stacking regions.

The reduction of bandwidth 
and related emergence of flat bands
identifies weakly doped MoS$_2$ bilayers as good candidates for
the observation of strong correlation effects. 
For a complete theoretical study of electronic correlations in these complex systems, it is important to take into account the atomic relaxation. 
We offer here a simple out-of-plane atomic displacements model for not too small rotation angles, typically a few degrees.
Preliminary investigations (not shown here) indicate
that  for determining the electronic structure of low-energy bands, the Slater-Koster tight-binding models \cite{Venky20,Zhan20}, that are efficient for rigidly twisted bilayers, would require a further adjustment of the parameters in order to be applicable to the $z$-modulated structures.

\section{Acknowledgments}

Calculations have been performed at the {\it Centre de Calcul (CDC)}, CY Cergy Paris Universit\'e,
and at GENCI-IDRIS (Grant
No. A0060910784). 
We thank Y. Costes and B. Mary, CDC, for computing assistance. 
This work was supported by the ANR project FlatMoi (ANR-21-CE30-0029)
and the Paris//Seine excellence initiative (Grants No. 2017-231-C01-A0 and AAP2019-0000000113).

\section{Authors contributions}
S. Venkateswarlu, A. Misssaoui, and G. Trambly de Laissardi\`ere performed the DFT calculations and the numerical analysis. 
S. Venkateswarlu, A. Honecker, and G. Trambly de Laissardi\`ere wrote and revised the manuscript. 
All authors discussed the results and approved the final version of the manuscript.

%
\bibliographystyle{ejp}
\bibliography{biblio}

\end{document}